\documentstyle[12pt]{article}
\begin{document}

\title{Perturbative QCD estimation of the
$B \rightarrow K^{*} + \gamma$ branching ratio}
\author{Marina-Aura Dariescu\thanks
{On leave of absence from the Dept. of Theoretical Physics,
^^ ^^ Al.I. Cuza'' University, 6600 Iasi, Romania}
\hspace{0.2cm} and \hspace{0.2cm}
C. Dariescu\thanks{MONBUSHO fellow, on leave of absence from
The Dept. of Theoretical Physics, ^^ ^^ Al.I. Cuza'' University,
 6600 Iasi, Romania} \\ \\ {\it Department of Physics} \\
{\it Toyama University} \\
{\it Gofuku, 930 Toyama, Japan} }
\date{}
\maketitle

\begin{abstract}
Working in a perturbative QCD model approach,
we obtain the essential form factor of the radiative
transition $B \rightarrow V^{*} \gamma$ and estimate
the branching ratio $BR(B \rightarrow K^{*} \gamma)$.
The results are determined by a parameter expressing
the momentum distribution in the $B$ - meson wave function.
Our estimations are compared to other theoretical
predictions as well as to experimental data.
\end{abstract}

\newpage

\section{Introduction}

Studies of rare $B$ decays are very active now
especially after CLEO II experiment reported the first
direct evidence for one-loop flavor-changing neutral
current diagram in
$B \rightarrow K^{*}(892) \gamma$ \cite{Cle:phy}.
The rare decays amplitudes proportional to
$V_{ub}$ as well as the so-called ^^ ^^ penguin''
processes are important for CP-violation measurements.
As $b \rightarrow u$ is suppressed by the corresponding
small value of CKM matrix element, the electromagnetic
$b \rightarrow s$ transition brings the dominant
contribution. Attempts in extending perturbative
QCD effects in factorized diagrams dominated by
a single gluon exchange with the spectator quark
developed by Szczepaniak, Henley and Brodsky
\cite{Szc:phy}, have been made in calculating
both semileptonic and nonleptonic $B$ - decays
with $0^{-}$ and $1^{-}$ mesons in final state \cite{Dar:rom,Dar:rjp}.
A comprehensive study of radiative rare decays,
using the spin symmetry for heavy quarks combined
with meson wave function models belong to Ali and
Greub \cite{Ali:phy} as well as to Ali, Ohl
and Mannel \cite{Ali:plb}.
Considerations on radiative transitions of $B$ mesons,
with an up to date review of experimental results
can be found in \cite{Bro:pre},
the theoretical implications of
$B \rightarrow K^{*} \gamma$ for the physics
beyond the standard model being also pointed out.

Our work is organized as follows.
After a brief review of the Szczepaniak,
Henley and Brodsky model, which is basic
for the rest of the material, in section 3
we put the amplitude of the radiative transition
$B \rightarrow V^{*} \gamma$ in terms of form factors.
In section 4 we start from the effective
hamiltonian
\begin{equation}
H = - \, \frac{G_{F}}{\sqrt{2}} \,
V_{tb} V_{ts}^{*} \,
C_{7}(m_{b}) \, \tilde{\cal{O}}_{7}(m_{B}) ,
\end{equation}
where $C_{7}(m_{b})$ containing the effects of
QCD corrections is the Wilson coefficient at
$\mu=m_{b}$ and $\tilde{\cal{O}}_{7}(m_{B})$
is computed in Brodsky and Lepage perturbative
QCD approach, replacing the usual one [6].
Than, the branching ratio of the two body decay
$B \rightarrow K^{*} \gamma$, expressed as a
function of the $a$ - parameter characterizing
the momentum distribution between the pair of
quarks in the  heavy $B$ meson is calculated.
Finally, we perform a comparison with other
theoretical results and with the experimental limits.

\section{Brief Review of the Model}

Our calculations being done within the framework
of perturbative QCD effects studied by Szczepaniak,
Henley and Brodsky \cite{Szc:phy}, it is worthwhile
to summarize here their approach in estimating
hadronic matrix elements of the transition currents.
Assuming the factorization for the case of an
exclusive nonleptonic decay of a heavy meson into
two much lighter $0^{-}$ mesons and neglecting
the final state interactions, they have written
the corresponding decay amplitude as a convolution
of a collinear irreducible hard-scattering quark-gluon
amplitude $T^{\mu}$ and the mesons wave functions
$\phi$ which, besides the spin factors \cite{Fie:nuc},
contain the fractional longitudinal momentum
distribution of the collinear quarks.
Distinct from the usual HQET \cite{Isg:nuc,Bjo:pre}
the dominant contribution in
$\alpha_{s}(Q^{2} \sim M_{B}^{2})$
is controlled by a single gluon exchange with the spectator.
After factorization and neglecting the disconnected
diagram whose contribution vanishes,
the matrix element of the transition current
between the two remaining hadronic states is put
in terms of form factors.
For a massive initial state, the decay amplitude is
of order $\alpha_{s}(Q^{2}) \sim 0.38 \,
(\Lambda_{QCD}^{2} = 0.01 \, {\rm GeV}^{2})$,
even without including loop corrections.

Although this method has been developed for exclusive
nonleptonic decays, we have applied it for both
nonleptonic and semileptonic heavy meson decays,
 with $1^{-}$ and $0^{-}$ light mesons in the final
state \cite{Dar:rom,Dar:rjp} and the obtained results,
compared with other theoretical predictions and
experimental upper limits, have encouraged us to test
its validity in estimating other decays,
such as the radiative transitions.
The first observation of the electromagnetic decay
$B \rightarrow K^{*} \gamma$, reported in 1993 by
CLEO II  \cite{Cle:phy}, has made this process of a
real theoretical interest, due to its implications
in the physics beyond the standard model \cite{Bro:pre}.
  Even the process is described by the one-loop
flavor-changing neutral current diagram,
it has been used factorization to estimate its
branching ratio, although it is not decided whether
factorization is a correct framework for the so-called
penguin diagrams. However, in this case one has to take
into account the short distance QCD corrections.

\section{Radiative Transition Form Factors}

The method described allows the calculation of the
heavy-to-light radiative transition
$B \rightarrow V^{*} \gamma$ in the same way as
in the nonleptonic decays, by considering a single
gluon exchange with the spectator (see fig.1). \\
\marginpar{Figure 1}
\vspace*{0.5cm}\\
To first order in $\alpha_{s}$, the matrix element of
the transition current can be computed as
\begin{eqnarray}
{T_{\mu}} = g_{s}^{2} \left\lbrace {\rm Tr}\left[
{\bar{\phi}}_{V} \sigma_{\mu \nu} (1- \gamma_{5})
q^{\nu} \frac{\gamma \cdot k_{B}+ m_{B}}{k_{B}^{2} -
m_{B}^{2}} \, \gamma_{\alpha} \, \phi_{B} \,
\gamma^{\alpha} \, \frac{\lambda_{a} \lambda^{a}}
{Q^{2}}\right] + \right. \nonumber \\ +
\left. {\rm Tr}\left[{\bar{\phi}}_{V} \gamma_{\alpha}
\frac{\gamma \cdot {k}_{V}}{k_{V}^{2}} \sigma_{\mu \nu}
(1- \gamma_{5}) q^{\nu}\phi_{B} \, \gamma^{\alpha}
\, \frac{\lambda_{a} \lambda^{a}}{Q^{2}}\right]
\right\rbrace ,
\end{eqnarray}
$\gamma \cdot k \; {\rm being} \; \gamma^{\beta}
k_{\beta}$. \\
Here,
\begin{equation}
\phi_{B}(x) = \varphi_{B} (x)
(\gamma \cdot P_{B} + m_{B})\gamma_{5}
\end{equation}
is the wave function of the heavy $B$ - meson
with the mass $m_{B}$ and the distribution
amplitude \cite{Szc:phy,Dar:rom,Dar:rjp}
\begin{equation}
\varphi_{B}(x) = \frac{f_{B}}{12} \,
\frac{x^{2} (1-x)^{2}}{[a^{2} x +(1-x)^{2}]^{2}} \cdot
\left\lbrace \int_{0}^{1} \frac{x^{2}(1-x)^{2}}
{[a^{2} x+(1-x)^{2}]^{2}} \; dx
\right\rbrace^{-1}
\end{equation}
where $a \sim 0.05-0.1$ is related to the maximum
of $\varphi_{B}(x)$ in the $B$ - meson,
and
\begin{equation}
\phi_{V} = \varphi_{V}(y) (\gamma \cdot P_{V}) \,
(\gamma \cdot \varepsilon)
\end{equation}
with
\begin{equation}
\varphi_{V}(y) = \frac{f_{V}}{12} \, y(1-y)
\left \lbrace\int_{0}^{1} y(1-y) \; dy
\right \rbrace^{-1}
\end{equation}
denotes the wave function of the light vector meson
of polarization vector $\varepsilon^{\mu}$.
The Tr means trace over spin, flavor and color indices
and integration over momentum fractions.

We put (2) in the form
\begin{equation}
{T_{\mu}} = i \varepsilon_{\mu \nu \alpha \beta}
\varepsilon^{\nu} q^{\alpha} P_{V}^{\beta}
V + \frac{\varepsilon_{\mu}}{2} m_{B}^{2}  A_{1} -
(\varepsilon \cdot q)(P_{V})_{\mu} A_{2} -
\frac{(\varepsilon \cdot q)}{2} q_{\mu} A_{3}
\end{equation}
pointing out the $a$ - dependent expression
of the essential form factor as being:
\begin{equation}
\mid V \mid = \mid A_{1} \mid = \mid A_{2} \mid =
64 \frac{g_{s}^{2}}{m_{B}^{2}} \left\lbrace \int_{0}^{1}
\frac{\varphi_{B}(x)}{1-x} dx \int_{0}^{1-a}
\frac{\varphi_{V}(y) (2-y)}{(1-y)^{2}} dy \right\rbrace ,
\end{equation}
where we have kept only the first order in
$1-x \sim a$.

For $B \rightarrow K^{*} \gamma$ the numerical values
of the form factor are from 0.25 (for $a=0.1$) to
0.6 (for $a=0.05$).

\section{Branching Ratio $BR(B \rightarrow K^{*} + \gamma)$}

In order to make an estimation for the exclusive
$B \rightarrow K^{*} \gamma$ branching ratio
\begin{equation}
BR(B \rightarrow K^{*} \gamma) = \frac{\Gamma
(B \rightarrow K^{*} \gamma)}{\Gamma_{tot}},
\end{equation}
we start from the effective hamiltonian (1)
with the operator $\tilde{\cal{O}}_{7}$ of the form
\begin{equation}
\tilde{\cal{O}}_{7} = \frac{e}{16 \pi^{2}} \, m_{B}
\, \eta^{\mu} \,T_{\mu},
\end{equation}
where $\eta^{\mu}$ is the polarization vector of the
photon and $T_{\mu}$ is (2), and we get for
$\Gamma(B \rightarrow K^{*} \gamma)$ the expression:
\begin{equation}
\Gamma(B \rightarrow K^{*} \gamma) =
\frac{\alpha G_{F}^{2} m_{B}^{5}}{128 \pi^{4}} \,
\mid V_{tb} V_{ts}^{*} \mid^{2}
\, \mid C_{7}(m_{b}) \mid^{2} \, \mid V \mid^{2}
\end{equation}
In order to numerically estimate the
$BR(B \rightarrow K^{*} \gamma)$ we take as input
the following numerical values \cite{Ali:phy}:
\begin{eqnarray}
& r_{u} \approx 7, \; r_{c} \approx 3, \;
\mid V_{ub} \mid \approx 0.0075, \;
\mid V_{cb} \mid = \mid V_{ts} \mid \approx 0.045,
& \nonumber \\ & G_{F} = 1.16637 \times 10^{-5}
{\rm GeV^{-2}} & \nonumber \\
& W_{eff}^{2} \simeq M^{2} - 2 M \cdot p_{F} \cdot 1.13,
{\rm with \; the \; parameter \;} p_{F} \approx 0.3 \,
{\rm GeV} &
\end{eqnarray}
in the $b$ quark width
\begin{equation}
\Gamma_{tot} = \frac{W_{eff}^{5} G_{F}^{2}}
{192 \pi^{3}} ( r_{u} \mid V_{ub} \mid^{2} +
r_{c} \mid V_{cb} \mid^{2} )
\end{equation}
Thus, we come to an $a$ - dependent branching ratio
whose values are from around $\simeq 0.59 \times 10^{-5}$
(for $a=0.1$) to $\simeq 3.43 \times 10^{-5}$ (for $a=0.05$).

A comparison to other theoretical models predictions and
to experimental data restricts the range of this parameter.
Thus, the estimation
$BR(B \rightarrow K^{*}(892) \gamma) =
(1.4-4.9) \times 10^{-5}$ [6] corresponds to the
$a$ - parameter in the range $a = 0.072 - 0.043$,
while a comparison to the branching ratio reported by
CLEO II \cite{Bro:pre} sets the upper limit for $a$
at about 0.06.

Our calculations have been made in the assumption
that the mass of $K^{*}$ can be neglected in comparison
to the mass of the $B$ meson. As mentioned in [2],
this has brought considerable simplifications.
Of course, for comparing the branching ratios for
different $K^{*}$ mesons one has to take into account
besides $\tilde{O}_{7}$, a second operator
$\tilde{O}_{7}^{'}$ proportional to $m_{K}$ and
to replace the wave function (5-6) with a type (3-4)
one, but with different $a$.
Thus, the branching ratios will depend on three parameters,
namely $a, a'$ and the mass parameter $z=m_{K}/m_{B}$.
However, testing whether Brodsky and Lepage approach
is applicable in these assumptions is beyond the aim
of the present paper.

\section{Conclusions}

We have developed an extension of our earlier
applications \cite{Dar:rom,Dar:rjp} of the Szczepaniak,
Henley and Brodsky procedure of perturbative QCD calculations
of hadronic current matrix element \cite{Szc:phy}
to the case of the rare
$B$ decay $B \rightarrow K^{*} \gamma$.
The branching ratio of this process is shown as a
function of the parameter $a$, having numerical values
from around $\simeq 0.59 \times 10^{-5}$ to
$\simeq 3.43 \times 10^{-5}$ for $a \in [0.05, 0.1]$,
the other theoretical model estimation [6]
corresponding to $a \leq 0.072$, while the experimental
lower limit of about $2.1 \times 10^{-5}$ imposing
$a \leq 0.06$. \\ \\

{\large{\bf Acknowledgments}} \\ \\
We would like to thank the anonymous referee
for very helpful observations. (C.D.) wishes
to express his gratitude to the Japanese Government
for financially supporting his work under a MONBUSHO
Fellowship. (M.A.D.) thanks Prof. D. Wyler and Dr. H. Simma
for having introduced her in Brodsky-Lepage approach.
We are also indebted to the Quantum Theory Group from
the Physics Department of Toyama University for their
kind hospitality and especially to Prof. T. Kurimoto
for suggesting us to tackle this subject.

\newpage
\begin{figure}

\vspace{5cm}
\caption{The contributing diagrams in the
hard scattering amplitude $T_{\mu}$}
\vspace{2cm}
\end{figure}

\end{document}